\documentclass[aps,prl,preprint,tightenlines,superscriptaddress,showpacs,byrevtex]{revtex4}

\usepackage{graphicx} 
\usepackage{dcolumn}  
\usepackage{amsmath, amsthm, amssymb} 

\graphicspath{{ps/}}

\newcommand{\ra}{\rightarrow}
\newcommand{\lam}{\Lambda}
\newcommand{\ks}{K^0_S}
\newcommand{\kp}{K^+}
\newcommand{\km}{K^-}
\newcommand{\dst}{D^{\ast}}
\newcommand{\dstm}{D^{\ast -}}
\newcommand{\dnb}{\overline{D}{}^0}
\newcommand{\lst}{\Lambda(1520)}
\newcommand{\tht}{\Theta(1540)^+}

\newcommand{\pfe}{|\vec{p}_F|}
\newcommand{\pk}{|\vec{p}_{\kp}|}
\newcommand{\ppk}{|\vec{p}_{pK}|}

\newcommand{\mev}{\mathrm{MeV}}

\newcommand{\mevc}{\mathrm{MeV}/c}
\newcommand{\gevc}{\mathrm{GeV}/c}
\newcommand{\mevm}{\mathrm{MeV}/c^2}
\newcommand{\gevm}{\mathrm{GeV}/c^2}

\newcommand{\br}{\mathcal{B}}
\newcommand{\cm}{\mathrm{cm}}

\begin{document}

\preprint{\vbox{ \hbox{   }
                 \hbox{BELLE-CONF-0518}
                 \hbox{LP2005-153}
                 \hbox{EPS05-490} 
}}
\vspace*{-2.0cm}
\begin{flushleft}
 \resizebox{!}{2.5cm}{\includegraphics{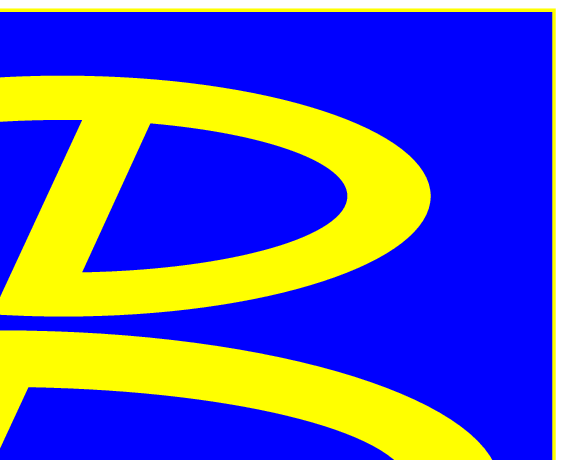}}
\end{flushleft}
\vspace*{-1.0cm}

\title{\boldmath\quad\\[0.5cm]
  Search for the $\tht$ pentaquark using kaon secondary interactions at Belle
}

\affiliation{Aomori University, Aomori}
\affiliation{Budker Institute of Nuclear Physics, Novosibirsk}
\affiliation{Chiba University, Chiba}
\affiliation{Chonnam National University, Kwangju}
\affiliation{University of Cincinnati, Cincinnati, Ohio 45221}
\affiliation{University of Frankfurt, Frankfurt}
\affiliation{Gyeongsang National University, Chinju}
\affiliation{University of Hawaii, Honolulu, Hawaii 96822}
\affiliation{High Energy Accelerator Research Organization (KEK), Tsukuba}
\affiliation{Hiroshima Institute of Technology, Hiroshima}
\affiliation{Institute of High Energy Physics, Chinese Academy of Sciences, Beijing}
\affiliation{Institute of High Energy Physics, Vienna}
\affiliation{Institute for Theoretical and Experimental Physics, Moscow}
\affiliation{J. Stefan Institute, Ljubljana}
\affiliation{Kanagawa University, Yokohama}
\affiliation{Korea University, Seoul}
\affiliation{Kyoto University, Kyoto}
\affiliation{Kyungpook National University, Taegu}
\affiliation{Swiss Federal Institute of Technology of Lausanne, EPFL, Lausanne}
\affiliation{University of Ljubljana, Ljubljana}
\affiliation{University of Maribor, Maribor}
\affiliation{University of Melbourne, Victoria}
\affiliation{Nagoya University, Nagoya}
\affiliation{Nara Women's University, Nara}
\affiliation{National Central University, Chung-li}
\affiliation{National Kaohsiung Normal University, Kaohsiung}
\affiliation{National United University, Miao Li}
\affiliation{Department of Physics, National Taiwan University, Taipei}
\affiliation{H. Niewodniczanski Institute of Nuclear Physics, Krakow}
\affiliation{Nippon Dental University, Niigata}
\affiliation{Niigata University, Niigata}
\affiliation{Nova Gorica Polytechnic, Nova Gorica}
\affiliation{Osaka City University, Osaka}
\affiliation{Osaka University, Osaka}
\affiliation{Panjab University, Chandigarh}
\affiliation{Peking University, Beijing}
\affiliation{Princeton University, Princeton, New Jersey 08544}
\affiliation{RIKEN BNL Research Center, Upton, New York 11973}
\affiliation{Saga University, Saga}
\affiliation{University of Science and Technology of China, Hefei}
\affiliation{Seoul National University, Seoul}
\affiliation{Shinshu University, Nagano}
\affiliation{Sungkyunkwan University, Suwon}
\affiliation{University of Sydney, Sydney NSW}
\affiliation{Tata Institute of Fundamental Research, Bombay}
\affiliation{Toho University, Funabashi}
\affiliation{Tohoku Gakuin University, Tagajo}
\affiliation{Tohoku University, Sendai}
\affiliation{Department of Physics, University of Tokyo, Tokyo}
\affiliation{Tokyo Institute of Technology, Tokyo}
\affiliation{Tokyo Metropolitan University, Tokyo}
\affiliation{Tokyo University of Agriculture and Technology, Tokyo}
\affiliation{Toyama National College of Maritime Technology, Toyama}
\affiliation{University of Tsukuba, Tsukuba}
\affiliation{Utkal University, Bhubaneswer}
\affiliation{Virginia Polytechnic Institute and State University, Blacksburg, Virginia 24061}
\affiliation{Yonsei University, Seoul}
  \author{K.~Abe}\affiliation{High Energy Accelerator Research Organization (KEK), Tsukuba} 
  \author{K.~Abe}\affiliation{Tohoku Gakuin University, Tagajo} 
  \author{I.~Adachi}\affiliation{High Energy Accelerator Research Organization (KEK), Tsukuba} 
  \author{H.~Aihara}\affiliation{Department of Physics, University of Tokyo, Tokyo} 
  \author{K.~Aoki}\affiliation{Nagoya University, Nagoya} 
  \author{K.~Arinstein}\affiliation{Budker Institute of Nuclear Physics, Novosibirsk} 
  \author{Y.~Asano}\affiliation{University of Tsukuba, Tsukuba} 
  \author{T.~Aso}\affiliation{Toyama National College of Maritime Technology, Toyama} 
  \author{V.~Aulchenko}\affiliation{Budker Institute of Nuclear Physics, Novosibirsk} 
  \author{T.~Aushev}\affiliation{Institute for Theoretical and Experimental Physics, Moscow} 
% ...
\collaboration{The Belle Collaboration}
\noaffiliation

\begin{abstract}
Using kaon secondary interactions in the material of the Belle
detector, we search for both inclusive and exclusive production of the
$\tht$. We set an upper limit of 2.5\% at the 90\%~C.L. on the ratio
of the $\tht$ to $\lst$ inclusive production cross sections. 
We also search for the $\tht$ as an intermediate resonance in the
charge exchange reaction $\kp n \to p\ks$. An upper limit of
$\Gamma_{\Theta^+}<0.64\,\mev$ at the 90\%~C.L. at
$m_{\Theta^+}=1.539\,\mevm$ is set. 
These results are obtained from a $397\,\mathrm{fb}^{-1}$ data
sample collected with the Belle detector near the $\Upsilon(4S)$
resonance, at the KEKB asymmetric energy $e^+ e^-$ collider.
\end{abstract}

\maketitle

{\renewcommand{\thefootnote}{\fnsymbol{footnote}}}

\setcounter{footnote}{0}

\section{Introduction}

The observation of the $\tht$ pentaquark,
an exotic bound state with the quark content $uudd\bar{s}$,
is one of the most puzzling mysteries of recent years 
(see~\cite{hicks} for an experimental overview). 
The $\tht$ was first observed in exclusive reactions at low energy~\cite{hicks}. 
Later several groups reported observation in inclusive reactions
at higher energies~\cite{hicks}.
Conversely, other experiments at high energies do not see the $\tht$
pentaquark although
they do observe significantly larger yields of conventional hyperons
than seen at the experiments that observe the $\tht$. 
In order
to resolve this discrepancy it is frequently assumed that 
pentaquark production decreases rapidly with energy. 
A high statistics experiment at low energies is therefore important. 
In order to achieve this, Belle utilises the small fraction of tracks
that interact with the material of the inner part of the detector. 
These secondary interactions are used to search for pentaquarks. Particles
produced in $e^+e^-$ annihilation at Belle have quite low
momenta; the most probable kaon momentum is only $0.6\,\gevc$.

Results from two analyses are presented. 
In the first, we search for inclusive production of the $\tht$ via the 
$KN \to \tht X$, $\tht\to p\ks$ process, using the signal from inclusive
$\lst$ production as a reference.
In the second, we search for exclusive $\tht$ production 
in the charge exchange reaction $\kp n\to\tht\to p\ks$. 
For this search, the yield of charge exchange reactions is used as a
reference, allowing a direct comparison to the results of the DIANA
experiment~\cite{diana}.

\section{Detector and data set}

These studies are performed using a $357\,\mathrm{fb}^{-1}$ data
sample collected at the $\Upsilon(4S)$ resonance and
$40\,\mathrm{fb}^{-1}$ at an energy $60\,{\mathrm{MeV}}$ below the
resonance. The data were collected with the Belle
detector~\cite{BELLE_DETECTOR} at the KEKB asymmetric energy $e^+ e^-$
storage rings~\cite{KEKB}.

The Belle detector is a large-solid-angle magnetic spectrometer that
consists of a silicon vertex detector (SVD), a 50-layer
cylindrical drift chamber (CDC), an array of aerogel threshold
Cherenkov counters (ACC), a barrel-like array of time-of-flight
scintillation counters (TOF), and an array of CsI(Tl) crystals (ECL)
located inside a superconducting solenoidal coil that produces a 1.5\,T
magnetic field. An iron flux return located outside the coil is
instrumented to detect muons and $K_L$ mesons (KLM).
Two different inner detector configurations were used. 
For the first sample of $155\,\mathrm{fb}^{-1}$, a $2.0\,$cm radius
beampipe and a 3-layer silicon vertex detector (SVD1) were used;
for the second sample of $242\,\mathrm{fb}^{-1}$, a $1.5\,$cm radius
beampipe, and a four-layer silicon vertex detector (SVD2), and a
small-cell inner drift chamber were used~\cite{new_svd}. 

A GEANT~\cite{geant} based Monte-Carlo (MC) simulation is used to
model the production of secondary $pK$ pairs, and to determine the
detector resolution and acceptance. 

\section{\boldmath Selection of secondary $pK$ pairs}

The analyses are performed by identifying $p\km$, $p\kp$ and $p\ks$
produced at secondary vertices. 
Charged particle candidates are required to be positively identified
based on the CDC ($dE/dx$), TOF and ACC information. 
In addition, proton and charged kaon candidates are removed if they are 
consistent with being electrons based on ECL, CDC and ACC 
information. 
$\ks$ candidates are reconstructed from $\pi^+\pi^-$ pairs
that have masses within $\pm 10\,\mevm$ of the nominal
$\ks$ mass (3$\sigma$ window). 
Additional selection requirements are imposed on the quality of the $\ks$
vertex, 
on the impact parameters of the daughter tracks
and on the angle between the momentum and the direction 
from the interaction point (IP) to the vertex. 

\begin{figure}[tbh]
\centering
\begin{picture}(550,220)
\put(5,-10){\includegraphics[width=1.04\textwidth]{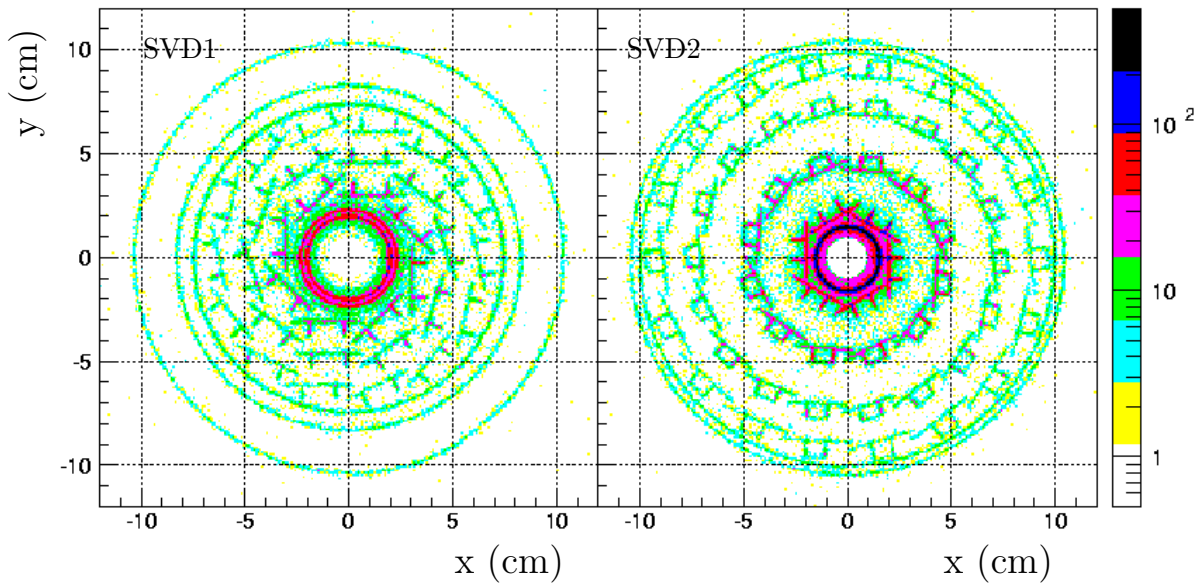}}
\put(11,170){\rotatebox{90}{\large y (cm)}}
\put(62,200){SVD1} 
\put(245,200){SVD2} 
\put(180,5){\large x (cm)} 
\put(365,5){\large x (cm)} 
\end{picture}
\caption{Distribution of reconstructed secondary $p\ks$ vertices in
  the Belle detector for the SVD1 (left) and SVD2 (right) data samples.}
\label{xy}
\end{figure}

The proton and kaon candidates are required to have an
origin that is displaced from the IP. The 
$p\km$, $p\kp$ and $p\ks$ vertices are fitted and 
those with a radial distance $1<R<11\,\cm$ are selected. 
Additional criteria on the quality of the $pK$ vertex
are applied.
We consider secondary $pK$ pairs only in the central part of the
detector $-0.74<\cos\theta<0.9$, where $\theta$ is the polar angle of
the secondary $pK$ vertex. 
The distributions of the secondary $p\ks$ vertices in the $xy$ plane
are shown in Fig.~\ref{xy} for the SVD1 and SVD2 data samples, 
where the $z$-axis passes through the IP and is antiparallel to the
$e^+$ beam.
The beam pipe, the SVD layers, the SVD cover and the CDC support
cylinders are clearly visible. 

\section{\boldmath Search for inclusive $\tht$ production}

\begin{figure}[tbh]
\centering
\begin{picture}(550,190)
\put(20,110){\rotatebox{90}{$\mathrm{N}/2\;\mevm$}} 
\put(230,110){\rotatebox{90}{$\mathrm{N}/0.2\;\gevc$}} 
\put(135,5){$m_{pK}\;(\gevm)$} 
\put(335,5){$p_{\lst}\;(\gevc)$} 
\put(180,160){(a)}
\put(390,160){(b)}
\put(20,-10){\includegraphics[width=8cm]{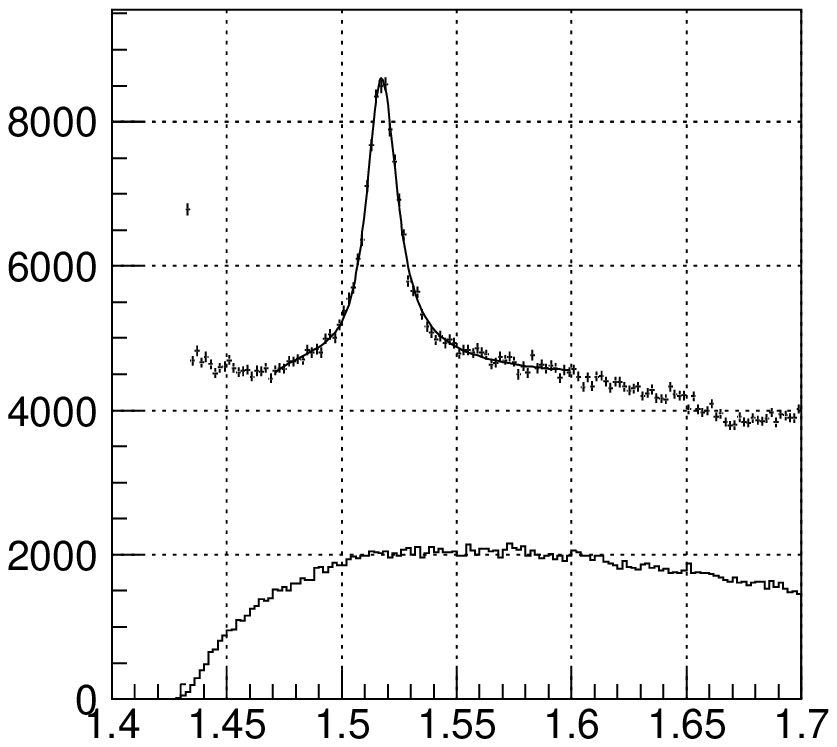}}
\put(230,-10){\includegraphics[width=8cm]{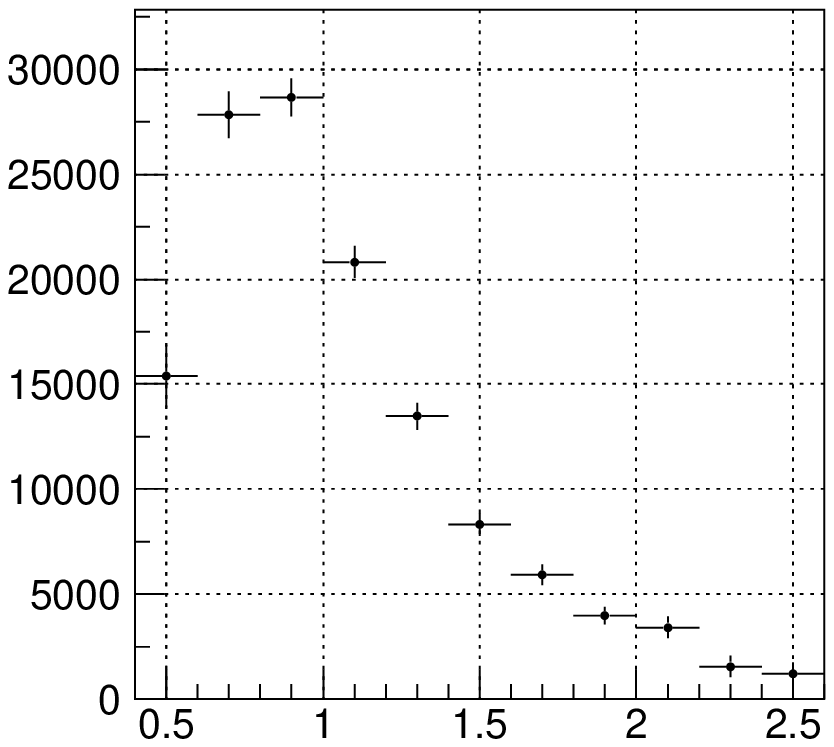}}
\end{picture}
\caption{(a) Mass spectra of $p\km$ (points with error bars) 
  and $p\ks$ (histogram) secondary pairs. The fit is described in the
  text. (b) Momentum spectrum of the $\lst$.
} 
\label{incl}
\end{figure}

The mass spectra for $p\km$ and $p\ks$ secondary vertices
are shown in Fig.~\ref{incl}. 
We apply an additional selection requirement on the angle between the $pK$
momentum and the direction from the IP to the $pK$ vertex $d\phi<1\,\mathrm{rad}$.
In the $p\km$ sample we reject $\lam\to p\pi^-$ and $\ks\to\pi^+\pi^-$
decays misidentified as secondary $p\km$ vertices. 
No significant structures are observed
in the $m_{p\ks}$ spectrum, 
while in the $m_{p\km}$ spectrum a $\lst$ signal is clearly visible. 
The $p\km$ mass spectrum is fitted to a sum of a $\lst$
signal function and a threshold function. 
The signal function is a $D$-wave Breit-Wigner shape 
convolved with a detector resolution function, determined from MC. 
The detector resolution function is parametrized by a double Gaussian
with widths of $2\,\mev$ and $5\,\mev$ with approximately equal
contributions from each Gaussian component.
The $\lst$ parameters obtained from the fit are consistent with the
PDG values~\cite{pdg}. 
The $\lst$ yield, defined as the fit function signal component
integrated over the $1.48$--$1.56\,\gevm$ mass interval ($2.5\Gamma$),
is $(4.02\pm 0.08)\cdot 10^4$ events. 

The $p\ks$ mass spectrum is fitted to a sum of a $\tht$
signal component and a third order polynomial. 
The $\tht$ is assumed to be narrow and its shape is 
determined by the detector resolution function, 
which is again a double Gaussian with similar parameters to those 
used for the $p\km$ mode.
The MC resolution is checked against data 
using the $\Omega^-\to\lam K^-$ signal, 
which has a topology similar to the secondary $p\ks$ pairs.
It is found that the $\Omega$ width is 4\% percent larger in data
than in MC, and this correction factor is applied to the mass
resolution of secondary $p\ks$ vertices. 
For $m=1540\,\mevm$ the fit yield is $58\pm 129$ events. 
Using the Feldman-Cousins method of upper limit
evaluation~\cite{feldman}, we find $N<270$~events at the 90\%~C.L.
The upper limit is below 320 events for a wide range of possible
$\tht$ masses. 
We set an upper limit on the ratio of $\tht$ to $\lst$ production 
cross sections:
\[
\frac{N_{\tht}}{N_{\lst}}
\frac{\epsilon_{p\km}}{\epsilon_{p\ks}}
\frac{\mathcal{B}(\lst\ra pK^-)}{\mathcal{B}(\tht\ra p\ks)
  \mathcal{B}(\ks\ra\pi^+\pi^-)}<2.5\% \;\text{at the 90\%~C.L.}
\]
It is assumed that $\mathcal{B}(\tht\ra p\ks)=25\%$. 
We take $\mathcal{B}(\lst\ra pK^-) = 
\frac{1}{2} \mathcal{B}(\lst\ra N\bar{K}) =
\frac{1}{2}(45\pm 1)\%$~\cite{pdg}.
The ratio of efficiencies for 
$\tht \ra p\ks$ and 
$\lst \ra p\km$ of about 41\% 
is obtained from MC simulation assuming that the two processes
have similar kinematics. 
Our limit is much lower than the results reported by many
experiments that observe the $\tht$. 
For example it is two orders of magnitude lower than the value
reported by the \mbox{HERMES} Collaboration~\cite{HERMES}.

The momentum spectrum of the produced $\lst$ is shown in
Fig.~\ref{incl}~(b). 
This spectrum is obtained by fitting $m_{p\km}$ in momentum
bins and correcting for the efficiency obtained from the MC. 
If a $\lst$ is produced as an intermediate resonance in the elastic
scattering of a $\km$ on a free proton, then its momentum is around 
$400\,\mevc$. As such, the observed hard momentum spectrum of the
$\lst$ confirms that they are produced in inelastic interactions.
We find that non-strange particles do not produce $\lst$ since the
secondary $p\km$ pairs are accompanied by an additional $\kp$ from
the same vertex in only about 0.5\% of events. 
The projectiles that can produce $\lst$ are 
$\km$, $\ks$, $K_L$ and $\lam$. 
The $\Lambda$ needs a momentum of about $1.8\,\gevc$ to produce
the $\lst$. The number of such $\Lambda$'s in $e^+e^-$ annihilations is
small. The observed number of $\lst$ can not be produced by $\Lambda$
projectiles even if the total $\Lambda N$ cross section
is assumed to be saturated by inclusive $\lst$ production.
We therefore conclude that our $\lst$ signal is due to kaon
interactions, with a contribution of $\lam$ interactions no larger
than a few percent.

\section{\boldmath Search for exclusive $\tht$ production}

Possible exclusive pentaquark production is studied using the
$\kp n\to p\ks$ reaction, searching for the $\tht$ as an intermediate
resonance. 
Since the projectile is not reconstructed, it is not possible to
distinguish this reaction from the reactions 
$\ks p\to p\ks$, $K_L^0 p\to p\ks$, and inelastic reactions with
a $\pi^0$ or undetected tracks in the final state. 
Therefore we apply selection criteria which suppress the contribution
of the inelastic reactions in the sample of secondary $p\ks$ pairs
(see section~\ref{pks_ul}), and determine the contribution of the
charge exchange reaction indirectly, as described below in this
section. 
The cross section of the $\ks p\to\tht\to p\ks$ reaction is expected
to be similar to the cross section of the $\kp n\to\tht\to p\ks$
reaction~\cite{kaidalov}, however the flux of $\ks$ is decreased by
decays in flight and we conservatively neglect the contribution of
this reaction. The $K_L^0 p\to p\ks$ reaction can produce a dip in the
$\tht$ region because of the interference of resonant and nonresonant
amplitudes. However, the size of the dip is expected to be
negligible~\cite{kaidalov}. 

The number of charge exchange reactions can be
estimated in a straightforward way using the known flux of primary
$\kp$, $\Phi^{\kp}$, the reaction cross section, $\sigma^\mathrm{ch}$, 
the amount of material, $M$, and the reconstruction efficiency 
for the secondary $p\ks$ pair, $\epsilon_{p\ks}$, and taking into
account nuclear effects:
\begin{equation}
N^\mathrm{ch}(m_{pK}) =  \int \Phi^{\kp} \; 
\sigma^\mathrm{ch} \; M \; \epsilon_{p\ks} \; \br \; 
S(E_N,\pfe)\; \delta(\sqrt{s} - m_{pK})\; 
P \; dE_N \; d^3p_F \; dp_{\kp} \; dR \; d\theta, 
\label{ch_long}
\end{equation}
where 
$\br$ is the product of $K^0$ branching fractions 
$\br = \br (K^0\to\ks)\,\br (\ks\to\pi^+\pi^-)$, 
$S(E_N,\pfe)$ is a nuclear spectral function  
which is a joint probability to find in a nucleus a nucleon with 
energy $E_N$ and Fermi momentum $\vec{p}_F$, 
$s=(E_{\kp}+E_N)^2-(\vec{p}_{\kp}+\vec{p}_F)^2$ is the centre of mass
(c.m.) energy of the reaction squared, 
$E_{\kp}$ is the energy and $\vec{p}_{\kp}$ is the momentum of the
projectile, 
$m_{pK}$ is the mass of the produced pair, 
$P$ is the probability that
the produced pair is not rescattered in the nucleus and 
$R$ and $\theta$ are the radial distance and polar angle of the 
secondary vertex, respectively.

However, it is difficult to accurately estimate the systematic
errors in this calculation because $M$ and $\epsilon_{p\ks}$
are complicated functions of the coordinates and the
estimation of $S$ and $P$ is model dependent. 
This problem is solved by reconstructing the decay chain 
$\dstm\ra\dnb\pi^-$, $\dnb\ra\kp\pi^-$ 
for events where a $\kp$ interacts elastically in the detector material. 
The reconstruction procedure is explained in detail in
section~\ref{ndst_proj}.
The yield of such decays, $N_{\dst}^{\mathrm{el}}$, is expressed as
\begin{equation}
N_{\dst}^\mathrm{el}(m_{pK}) =  \int \Phi^{\kp}_{\dst} \; 
\sigma^\mathrm{el} \; M \; \epsilon_{p\kp} \; 
S(E_N,\pfe)\; \delta(\sqrt{s} - m_{pK})\; 
P \; dE_N \; d^3p_F \; dp_{\kp} \; dR \; d\theta, 
\label{el_long}
\end{equation}
where $\Phi^{\kp}_{\dst}$ is the flux of $\kp$ originating from the
selected $\dstm$ decay,
$\sigma^\mathrm{el}$ is the cross section for elastic $\kp p\ra
p\kp$ scattering and 
$\epsilon_{p\kp}$ is the efficiency of reconstructing the secondary
$p\kp$ vertex. 
The nuclear suppression $P$ is assumed to be the same for $p\kp$ pairs
and $pK^0$ pairs produced in the charge exchange reaction. 
We express $N^\mathrm{ch}$ in terms of $N_{\dst}^{\mathrm{el}}$.
The expressions are simplified by the facts that 
the products $M \epsilon_{p\ks}$ and $M \epsilon_{p\kp}$ are
approximately independent of $\theta$ for the central part of the
detector;
the ratio $\epsilon_{p\ks}/\epsilon_{p\kp}$ is approximately
independent of the secondary pair momentum, $\ppk$; 
and the nuclear suppression $P$ is approximately 
independent of $\ppk$ (see section~\ref{ndst_proj}). 
We approximate the nuclear spectral function as
$S(E_N,\pfe)=W(\pfe)\;\delta(E_N-f(\pfe))$, where the function
$f(\pfe)$ is defined so that $E_N=f(\pfe)$ corresponds to the maximum
of $S(E_N,\pfe)$.
We obtain
\begin{equation}
N^\mathrm{ch}(m_{pK}) =  N_{\dst}^{\mathrm{el}}(m_{pK})  
\frac{\Phi^{\kp} (m_{pK})}{\Phi^{\kp}_{\dst} (m_{pK})} 
\frac{\sigma^\mathrm{ch} (m_{pK})}{\sigma^\mathrm{el} (m_{pK})} 
\frac{\epsilon_{p\ks}(m_{pK})\;\br}{\epsilon_{p\kp}(m_{pK})},
\label{main_formula}
\end{equation}
where
\begin{equation}
\frac{\Phi^{\kp} (m_{pK})}{\Phi^{\kp}_{\dst} (m_{pK})} = 
\frac{\int \Phi^{\kp} (\pk)\; W(\pfe)\; 
\delta(\sqrt{s} - m_{pK})\; 
\epsilon_{p\kp} (m_{pK},\ppk)\; d^3p_F \; dp_{\kp}}
{\int \Phi^{\kp}_{\dst} (\pk)\; W(\pfe)\; 
\delta(\sqrt{s} - m_{pK})\; 
\epsilon_{p\kp} (m_{pK},\ppk)\; d^3p_F \; dp_{\kp}}.
\label{raflu}
\end{equation}
The p.d.f. of the Fermi momentum distribution $W(\pfe)$ is determined 
from data in section~\ref{ndst_proj}. The form of the $f(\pfe)$ is
discussed in the same section. The integrations are described in
section~\ref{phi_prim}. 
Equation~\eqref{main_formula} provides the basic formula to determine
the yield of the charge exchange reaction in our data sample.
The formula~\eqref{main_formula} would be even simpler if we could
use $\kp$ projectiles from $\dstm$ decays that undergo charge
exchange in the detector material. However, the fraction of such 
events is very low because of the relatively small cross section, and
the low $\ks$ reconstruction efficiency and branching fraction. 

\subsection{Determination of $N_{\dst}^{\mathrm{el}}$}
\label{ndst_proj}

To determine the number of $\dstm\ra\dnb\pi^-$,
$\dnb\ra\kp\pi^-$ decays for which a $\kp$ interacts elastically
in the detector material,
the four-momentum of the interacting $\kp$ is reconstructed based on
the information available from the produced secondary $p\kp$ pair. 

If a secondary $pK$ pair is produced in a quasi-elastic reaction, 
four-momentum conservation implies
\begin{align}
E_K + E_N = E_{pK}, \label{e_cons} \\
\vec{p}_K + \vec{p}_F = \vec{p}_{pK}. \label{p3_cons}
\end{align}
The nucleon energy, $E_N$ is approximated by~\cite{sibirtsev}
\begin{equation}
E_N = m_N - 2\epsilon - \frac{\vec{p}_F^{\; 2}}{2m_N},
\label{en_sibir}
\end{equation}
where $m_N$ is the nucleon mass and $\epsilon\sim 7\,\mev$ is the nucleon
binding energy. This approximation is valid for the high $\pfe$ part of
the nuclear spectral function. 
Another possible approximation is 
$E_N = m_N - E_R$, where $E_R\sim 26\,\mev$ is the average
removal energy of the bound nucleon. 
We find that both approximations give a similar resolution in the
$\dnb$ mass (see below) and our result is independent of the choice. 
The quantities $E_{pK}$ and $\vec{p}_{pK}$ are measured; 
taking into account the primary and secondary vertex constraints, 
Eqs~\eqref{e_cons}--\eqref{en_sibir} can then be solved iteratively. 

The iterative process is started from Eq.~\eqref{en_sibir} where some
average value of $\pfe$ is substituted. Then the energy of the
projectile $E_K$ is determined from Eq.~\eqref{e_cons}.
In the next step the projectile momentum $\vec{p}_K$ is determined
from its absolute value 
$|\vec{p}_K|=\sqrt{E_K^2- m_K^2}$, and the flight direction
obtained from the primary and secondary vertex constraints,
taking into account the bending of the track in the magnetic field. 
The value of $\pfe$ is then determined from Eq.~\eqref{p3_cons} and
the iteration loop is closed by substituting the obtained $\pfe$ into 
Eq.~\eqref{en_sibir}. 

\begin{figure}[tbh]
\centering
\begin{picture}(550,190)
\put(20,110){\rotatebox{90}{$\mathrm{N}/10\;\mevm$}}
\put(230,110){\rotatebox{90}{$\mathrm{N}/10\;\mevc$}} 
\put(110,5){$m_{\kp_\mathrm{prim}\pi^-}\;(\gevm)$} 
\put(350,5){$\pfe\;(\gevc)$} 
\put(180,160){(a)}
\put(390,160){(b)}
\put(20,-10){\includegraphics[width=8cm]{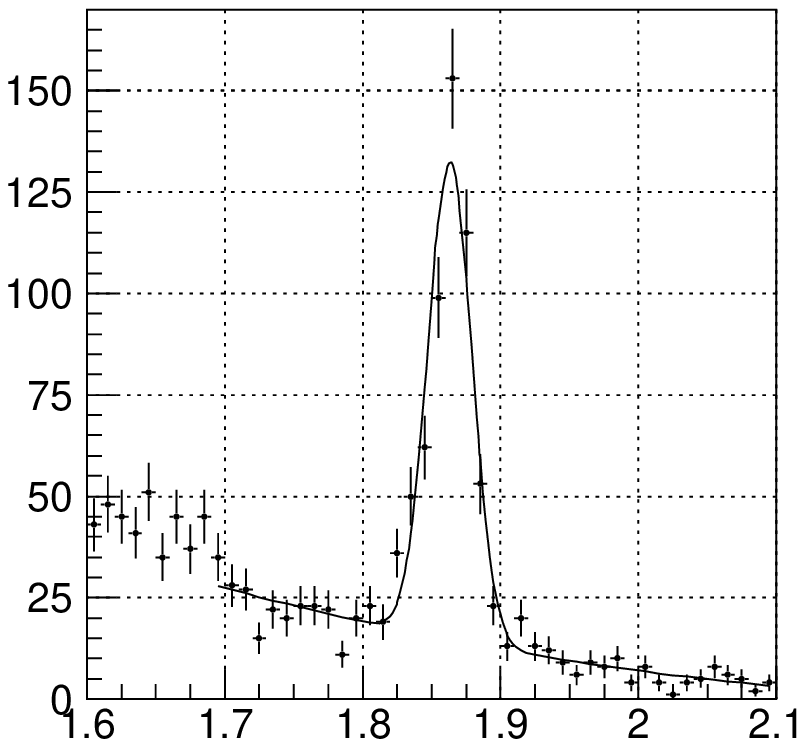}}
\put(230,-10){\includegraphics[width=8cm]{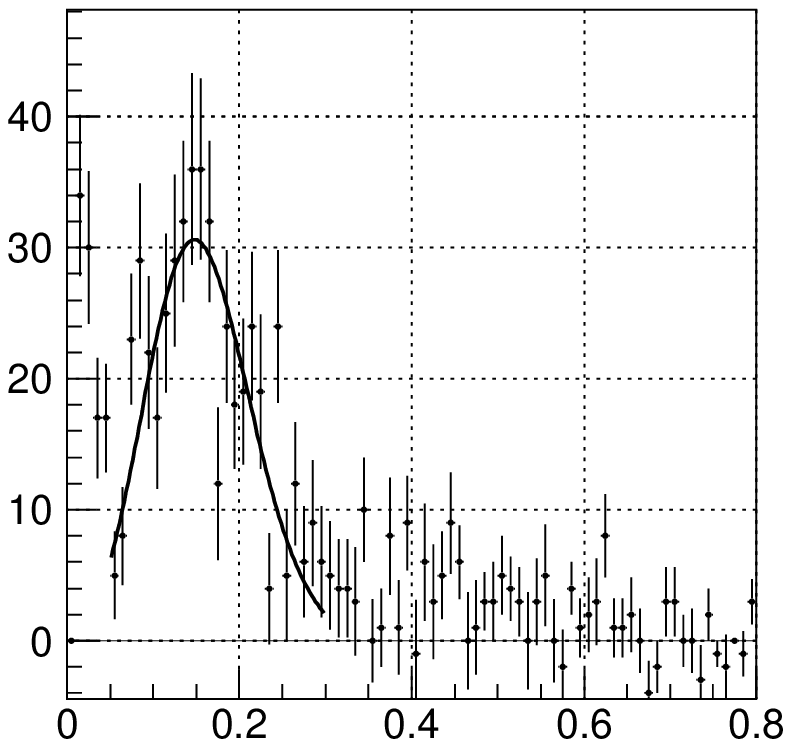}}
\end{picture}
\caption{(a) Invariant mass of the selected $\dnb$ candidates with
  the $\kp$ reconstructed via a secondary $p\kp$ vertex.
  (b) Sideband subtracted distribution of $\pfe$ for the selected
  $\dnb$ candidates, fitted to the oscillator model expectations.} 
\label{dn_sec}
\end{figure}

The projectile four-momentum is determined for all secondary $p\kp$
pairs. The resulting $\kp$ projectile candidates are then combined
with all the $\pi^-$ candidates in the event to form $\dnb$
candidates; the $\dnb$ candidates are combined with all the remaining
$\pi^-$ candidates to form $\dstm$ candidates. 
The $\pi^-$ candidates are required to be positively identified based
on the CDC ($dE/dx$), TOF and ACC information and to originate from
the vicinity of the IP. 
We reject vertices with additional tracks and require
$50<\pfe<300\,\mevc$. 
The lower bound on $\pfe$ is used to reject interactions on hydrogen
(see below). 
Events in a $\pm 2\,\mevm$ ($3\sigma$) window in $\Delta
m_{\dst}=m_{\dst}-m_{\dnb}$ are selected and the mass of the daughter
$\dnb$ candidates is plotted (see Fig.~\ref{dn_sec}~(a)).
A signal of  $470 \pm 26$ $\dnb$  events with the mass consistent with
the PDG value and a mass resolution of $16\,\mevm$ is observed.
The $\dnb$ signal and sideband regions are selected as 
$|m_{\kp\pi^-}-m_{\dnb}|<50\,\mevm$ and 
$60<|m_{\kp\pi^-}-m_{\dnb}|<110\,\mevm$, respectively. 
The sideband subtracted distribution of $\pfe$ is shown in
Fig.~\ref{dn_sec}~(b). 
The peak near zero is attributed to interactions on hydrogen,
which is present in the detector material. If $E_N = m_N$ is used
instead of Eq.~\eqref{en_sibir}, this peak is found at zero,
as expected for interactions with a free proton.
The $\pfe$ spectrum is fitted to the parametrization, expected in the
oscillator model~\cite{pf_distr}:
$\pfe^2[1+4/3(\pfe/p_0)^2]\exp(-\pfe^2/p_0^2)$. The value for the
model parameter returned by the fit, $p_0=115\pm 4\,\mevc$, 
is comparable to those obtained from other measurements of $\pfe$ 
distributions~\cite{pf_distr}. 

We find that the fraction of $\dstm$ events in the hydrogen peak is 
roughly independent of pair momentum. Since this fraction is
inversely proportional to $P$ 
(there can be no rescattering in hydrogen), 
we conclude that $P$ is also roughly independent of pair
momentum, which is important to perform the integration in
Eq.~\eqref{raflu}. 

The $m_{\dnb}$ spectra are fitted in $m_{pK}$ bins to determine the
$\dstm$ yield. The fit function is comprised of the sum of a Gaussian
and a first order polynomial. The Gaussian mean and width and the
polynomial parameters are all floated in the fit. The number of
$\dstm$ mesons in the $\tht$ region is $24\pm 7$ per $50\,\mevm$ bin.

\subsection{Determination of $\Phi^{\kp} / \Phi^{\kp}_{\dst}$}
\label{phi_prim}

The fluxes of primary $\kp$'s and $\kp$'s from $\dstm$ are 
determined from data. 
Primary $\kp$ candidates are required to originate from the vicinity
of the IP and to be positively identified based on the CDC ($dE/dx$),
TOF and ACC information.
$\dstm$ candidates are selected by combining the kaon candidate
with pion candidates, in the same way as described above. 
Correction for reconstruction efficiency and contamination from
other particle species is performed using MC that is calibrated
from data.
The integration in Eq.~\ref{raflu} is performed using a Monte Carlo
technique. 
The nucleon Fermi momentum $\vec{p}_F$ is assumed to be isotropic
relative to the projectile momentum $\vec{p}_K$. 
The fact that the secondary $pK$ pair efficiency depends on the
momentum results in a correction of about 3\% in the ratio.
The flux ratio at $m_{pK}=1.539\,\gevm$ is equal to $850\pm 20$; 
the uncertainty is dominated by the assumption concerning the relation
between $E_N$ and $\pfe$.
For this measurement we use about 20\% of the data sample distributed 
uniformly over the running period.

\subsection{Determination of $\sigma^\mathrm{ch} / \sigma^\mathrm{el}$}

The cross sections $\sigma^\mathrm{ch}$ and $\sigma^\mathrm{el}$ are
obtained from published data~\cite{ch,pdg}. 
The data are fitted with polynomials in $m_{pK}$, and the ratio of the
fitted functions is used to obtain
$\sigma^\mathrm{ch}/\sigma^\mathrm{el}$. 
The value of the ratio is $0.35\pm 0.02$ at $m_{pK}=1.539\,\gevm$ and
rises with $m_{pK}$. Errors are assigned based on the typical
experimental errors in the region of interest.

\subsection{Determination of $\epsilon_{p\ks} / \epsilon_{p\kp}$}

Monte Carlo simulations are used to estimate the ratio
$\epsilon_{p\ks}/\epsilon_{p\kp}$.  
The angular distribution in the reaction c.m. frame is assumed to be
uniform as expected for low energy elastic $\kp p$ scattering
and for $\tht\to p\ks$ decay. 
We consider the following sources of systematic uncertainty:
$\ks$, $\kp$ and secondary $pK$ reconstruction efficiency (7\%), 
uncertainty in material description ($5\%$), 
uncertainty in the description of the reaction kinematics (5\%), 
and the MC statistical uncertainty ($5\%$).
The ratio of efficiencies at $m_{pK}=1.539\,\gevm$ is $(43\pm 5)\%$. Here all
the uncertainties are added in quadrature.

\subsection{Upper limit on $\tht$ yield in exclusive reaction}
\label{pks_ul}

\begin{figure}[tbh]
\centering
\begin{picture}(550,190)
\put(20,110){\rotatebox{90}{$\mathrm{N} / 2\,\mevm$}}
\put(230,130){\rotatebox{90}{$\Gamma,\mev$}} 
\put(130,5){$m_{p\ks}\;(\gevm)$} 
\put(340,5){$m_{p\ks}\;(\gevm)$} 
\put(65,160){(a)}
\put(390,160){(b)}
\put(20,-10){\includegraphics[width=8cm]{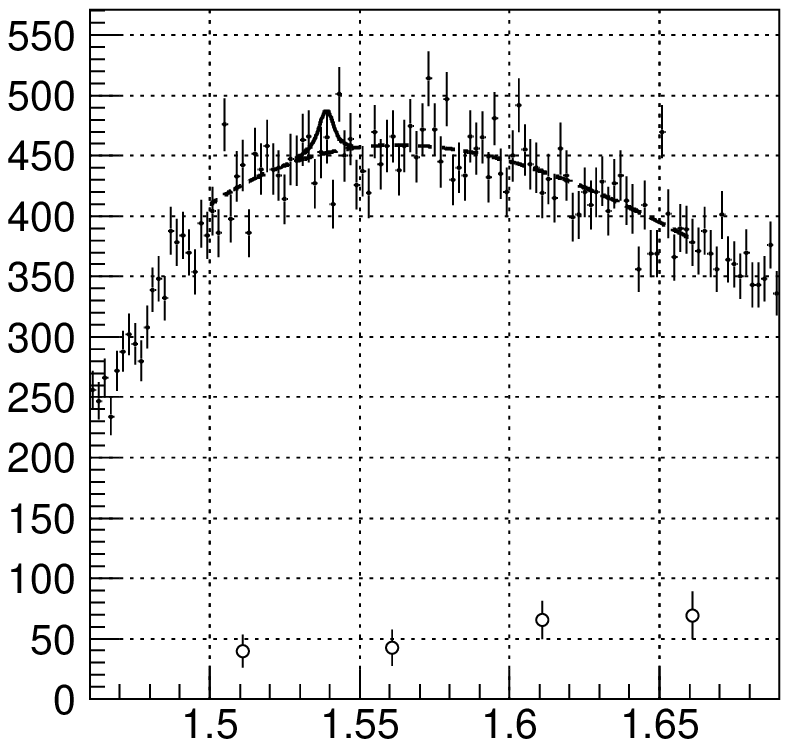}}
\put(230,-10){\includegraphics[width=8cm]{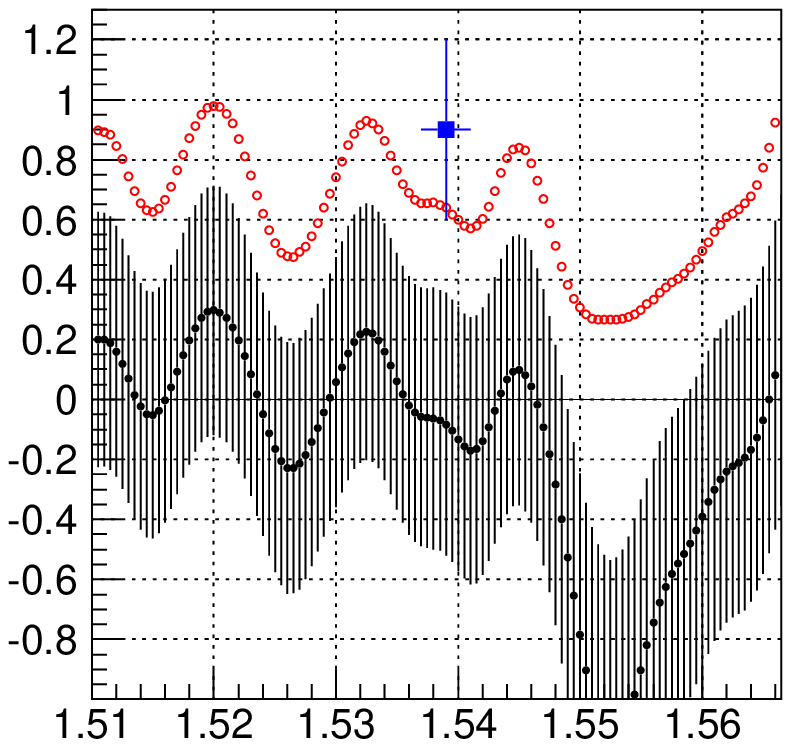}}
\end{picture}
\caption{(a) Invariant mass spectrum for secondary $p\ks$ pairs
  (small dots with error bars) and expected yield of the charge exchange reaction per
  $2\,\mevm$ (open dots). 
  A fit to a third order polynomial is shown with a dashed curve. 
  The $\tht$ contribution expected from the DIANA result is
  shown with solid line. 
  (b) The yield of $\tht$ from the fit, expressed in terms of the
  resonance width. The open dots correspond to the
  upper limit at the 90\%~C.L., obtained using the Feldman-Cousins method. The
  square with error bars indicates the current PDG value for the $\tht$
  width.
} 
\label{excl}
\end{figure}

To suppress the contribution of inelastic reactions in the sample
of secondary $p\ks$ pairs we reject vertices with additional
tracks and require $50<\pfe<300\,\mevc$. 
The lower bound on $\pfe$ is used to reject interactions on hydrogen,
which do not contribute to the charge exchange reaction.
The effect of angular cuts used by DIANA to suppress rescattering is
checked, and found not to suppress background significantly when the
$\pfe$ cut is applied, and not to improve the sensitivity.
The $p\ks$ mass spectrum and the expected yield from the
charge exchange reaction is shown in Fig.~\ref{excl}~(a). 
The statistical and systematic uncertainties on $N^\mathrm{ch}$ are
added in quadrature. 
In the $\tht$ region we expect $(1.03\pm 0.36)\cdot 10^3$
charge exchange events per $50\,\mevm$ bin. 
The $m_{p\ks}$ distribution is fitted to a third order
polynomial and a signal p.d.f. positioned at various values of
$m_{p\ks}$. The fit finds $N_{\Theta^+}=-11\pm 59$ candidates at
$m_{pK}=1.539\,\gevm$. 
The ratio of the $\tht$ yield to the charge exchange reaction yield
can be expressed in terms of the $\tht$ width 
(see for example~\cite{cahn_trilling}):
\[
\Gamma_{\Theta^+}=\frac{N_{\Theta^+}}{N^\mathrm{ch}}
\frac{\sigma^\mathrm{ch}}{107\,\mathrm{mb} \; B_i \; B_f} \; \Delta m,
\]
where $B_i$ and $B_f$ are $\tht$ branching fractions into initial and
final states, $B_i=B_f=0.5$, and $\Delta m$ is the mass interval of
$p\ks$ pairs used to determine $N^\mathrm{ch}$.
The resulting values of $\Gamma_{\Theta^+}$ are shown as a function of 
$m_{pK}$ in Fig.~\ref{excl}~(b).
Also shown in Fig.~\ref{excl}~(b) are the 90\% C.L. upper
limits, obtained with the Feldman-Cousins method~\cite{feldman}, 
and the current PDG value for the $\tht$ width~\cite{pdg}, which is 
based on the DIANA result. 
A similar width has been inferred from a reanalysis of $\kp d$
scattering~\cite{kpdeutron}. 
It is assumed that nuclear suppression for the $\tht$ is the
same as for nonresonant $p\ks$ pairs. 

\section{Conclusions}

Using kaon interactions in the material of the Belle detector, 
we searched for both inclusive and exclusive production of the $\tht$. 
No $\tht$ signal was found in the sample of secondary $p\ks$ pairs. 
For inclusive production we set the following upper limit:
\[
\frac{\sigma(KN \to \tht X)}{\sigma(KN \to \lst X)} < 2.5\% \;\text{at
  the 90\%~C.L.}
\]
For exclusive production we find
\[
\Gamma(\kp n\to\tht\to p\ks) < 0.64\,\mev \;\text{at the 90\%~C.L.}
\]
at $m_{\Theta^+}=1.539\,\mevm$. 
This upper limit is below the current PDG value of 
\mbox{$\Gamma=0.9\pm 0.3\,\mev$}, and below $1.0\,\mev$ 
for a wide interval of possible $\tht$ masses. 
This measurement uses a sample of low energy kaon interactions and
allows for a direct comparison with the DIANA result. With similar
sensitivity, our results do not support their evidence for the
$\tht$. 

\section{Acknowledgements}

We are grateful to A.~Kaidalov and Yu.~Kiselev for the fruitful
discussions. 
We thank the KEKB group for the excellent operation of the
accelerator, the KEK cryogenics group for the efficient
operation of the solenoid, and the KEK computer group and
the National Institute of Informatics for valuable computing
and Super-SINET network support. We acknowledge support from
the Ministry of Education, Culture, Sports, Science, and
Technology of Japan and the Japan Society for the Promotion
of Science; the Australian Research Council and the
Australian Department of Education, Science and Training;
the National Science Foundation of China under contract
No.~10175071; the Department of Science and Technology of
India; the BK21 program of the Ministry of Education of
Korea and the CHEP SRC program of the Korea Science and
Engineering Foundation; the Polish State Committee for
Scientific Research under contract No.~2P03B 01324; the
Ministry of Science and Technology of the Russian
Federation; the Ministry of Higher Education, Science and Technology
of the Republic of Slovenia;  the Swiss National Science Foundation;
the National Science Council and 
the Ministry of Education of Taiwan; and the U.S.\
Department of Energy.

\end{document}